# A systematic review of generative AI usage for IT project management

*Ionut Anghel and Tudor Cioara\**
*Computer Science Department, Technical*
*University of Cluj-Napoca,Cluj-Napoca, Romania*
*{ionut.anghel, tudor.cioara}@cs.utcluj.ro*

**Abstract:** This paper aims to synthesize current knowledge on generative AI in IT project management using the PRISMA methodology to provide researchers with a comprehensive perspective on techniques, applications, adoption trends, limitations, and integration across project management tools and process groups. The analysis reveals a clear dominance of OpenAI's GPT in the included studies but relying primarily on prompt engineering, suggesting that research in this area remains at an exploratory stage. Finally, it identifies and discusses three promising research directions for AI-enabled project management, including process group-specific AI agents, project role-based AI agents, and hybrid collaborative networks that enable human-guided orchestration.

## Introduction

Information Technology (IT) project management (PM) has become more challenging due to rapid technological change, pandemic impact on the relocation of work from office settings to virtual environments, and growing scale and complexity of data generated and processed by organizations. Traditional approaches rely on static planning tools, manual coordination, and human judgment struggling to handle dynamic and high-uncertainty environments that characterize software development, digital transformation projects, and IT infrastructure initiatives (Nenni et al., 2025). Even though substantial organizational investment has been made in IT projects, a significant proportion of them still fail to achieve their planned scope, time and cost targets (Abdulghafour et al., 2024). There is a need to develop new intelligent and adaptive techniques for supporting complex decision-making in IT project contexts, particularly as advances in Artificial Intelligence (AI) create new opportunities for data-driven project management (Khalil et al., 2025).

According to PMI (Project Management Institute), PM is typically structured around a lifecycle framework, which organizes projects into five distinct process groups (Rose, 2013): initiating, planning, executing, monitoring and controlling, and closing. Each process group involves distinct decision-making processes and managerial activities. During initiation, project objectives, scope, and stakeholder expectations are defined. Planning focuses on resource allocation, scheduling, budgeting, and risk assessment. Execution deals with task implementation and team coordination, while monitoring and controlling involve performance measurement, issue resolution, and adaptation to unforeseen events. Finally, closing deals with formal completion of project

*Corresponding author*

deliverables, evaluation of outcomes, and documentation of lessons learned. For IT projects, these process groups are often iterative and agile, involving sprint planning, continuous integration, and frequent stakeholder reviews, which increase both the complexity and volume of project data (Dong et al., 2024).

AI technologies can reshape IT PM by embedding intelligence into core activities, leading to more effective planning, executing, tracking, and control (Müller et al., 2024). Using Machine Learning (ML), Natural Language Processing (NLP), or advanced optimization algorithms, AI systems can extract insights from extensive datasets, predict future states, and reduce manual effort through automation (Alharthi et al., 2023) (Wang et al., 2025). In IT projects, AI applications can be integrated in each stage of the project lifecycle: aiding requirement analysis and risk forecasting during planning, automating monitoring and reporting in execution, and facilitating performance evaluation during closing (Almeida et al., 2025). Among AI technologies, Generative AI (GenAI), particularly transformer-based large language models (LLMs), can prove to be a powerful tool (Karnouskos, 2024). GenAI systems can generate code snippets, project documentation, test scripts, and project reports, by offloading repetitive activities, enabling IT specialists to concentrate on higher-value problem-solving and strategic choices (Lee et al, 2025). This new class of AI tools introduces both opportunities and challenges, necessitating careful attention to safety, governance, and data privacy issues in IT environments, where sensitive corporate, client, and employee data are frequently processed (Singh et al., 2025).

However, literature still lacks comprehensive systematic reviews that consolidate its use, benefits, limitations, and future research directions and implications across the PM lifecycle. Existing reviews are focused on rather general use of AI in PM mainly analyzing areas such as forecasting accuracy, risk mitigation and stakeholder collaboration (Salimimoghadam et al., 2025), (Khalil et al., 2025), (Alharthi et al., 2023). Only few recent studies are focused on the specific use case of GenAI in IT PM, and they only partially address this topic. (Assalaarachchi et al., 2025) focuses on the perspectives of software practitioners regarding the use of GenAI for automating routine tasks, predictive analytics, communication and collaboration, and agile practices, rather than providing a comprehensive synthesis of literature. (Aramali et al., 2025) concentrates mainly on tools such as ChatGPT in project management, examining their influence on corporate values, employee perceptions, and practical applications across project process groups and roles. While these studies provide useful insights, they do not offer a systematic overview of the broader literature on GenAI applications in IT project management or identify consolidated research directions. Comprehensive reviews are necessary to aggregate prior work, pinpoint limitations in literature, and systematically explain how GenAI can be effectively and responsibly applied in IT project contexts and can guide both researchers and practitioners to take advantage of GenAI's potential. Moreover, the continuous evolution and innovations for GenAI models (e.g. AI agents and agentic AI (Hughes et al., 2025)) requires new analysis that can explore the possible application of latest models in different PM process groups. Existing studies focus only on isolated tools, individual lifecycle process groups, leaving gaps in understanding the AI agent's adoption across platforms and methodologies.

This review synthesizes current knowledge on GenAI in IT PM, highlighting the most important techniques, new trends, frameworks for human and AI agents' collaboration, and commercial/open-source platforms, while linking AI applications to the Project management Body of Knowledge (PMBOK) process groups, also called focus areas in the newest eight edition (PMBOK Guide, 2026). Also, we identify and discuss three promising approaches for AI in project management that are emerging. The first involves process group specialized AI agents that provide targeted support for planning, execution, and reporting. The second focuses on AI agents that replicate and simulate project management roles, acting as virtual team members such as project managers, risk managers, and scrum masters. The third combines these strategies, with process group-specific agents informed by role-based simulations, creating a network of AI collaborators that can operate in parallel while human managers oversee final decisions. Together, these approaches aim to enhance decision-making, improve efficiency, and transform project management from reactive oversight to proactive orchestration, all while keeping humans actively in the loop.

The aim is to provide a comprehensive perspective for researchers and practitioners seeking to leverage GenAI technologies in IT project contexts. Our goal is to answer the next research questions:

- What are the current applications of GenAI across the IT PM lifecycle, and what are the key trends shaping its adoption and evolution in this domain?
- What are the main limitations, and open challenges associated with the integration of GenAI into IT PM practices and tools?
- To what extent are GenAI tools integrated into commercial and open-source PM platforms, and how do these integrations support different process groups of the project lifecycle?
- What emerging roles, capabilities, and governance challenges are associated with the adoption of AI agents in IT project management?

We conducted an extensive literature search using Web of Science (WoS) database to identify papers addressing the application of GenAI in IT PM. The search targeted key thematic areas including GenAI and LLMs integration across project process groups, prompt engineering and model adaptation. Inclusion criteria comprised journal articles indexed in WoS, written in English and published between 2021-2026, from leading publishers. Studies focusing exclusively on general AI applications without specific reference to GenAI or IT PM contexts were excluded, ensuring the review remained focused on the most relevant and recent advances in the field.

## From traditional AI to GenAI in PM

To ensure conceptual clarity, this section briefly describes the broader role of AI in project management, highlighting prior developments in predictive analytics, automation, and decision-support systems. This contextual overview helps in making the boundaries between traditional AI and Generative AI, thereby justifying the specific focus of the systematic review. The papers used in this section are illustrative and do belong to the formal review dataset.

Within the traditional AI landscape, supervised Machine Learning (ML) algorithms (e.g. Random Forests, Support Vector Machines, Gradient Boosting) have established themselves as reliable tools for forecasting project indicators such as delivery timelines, budget, and potential resource constraints (Alharthi et al., 2023). Deep learning architectures such as Convolutional Neural Networks (CNNs) and Long Short-Term Memory (LSTM) networks allow analysis of unstructured and temporal data, including code repositories, bug reports, and communication logs (Wang et al., 2025). Natural language processing (NLP) enables processing and analysis of requirements documents, emails, conversation logs, and stakeholder communications, improving documentation, sentiment analysis, and reporting (Su, 2025). Optimization algorithms such as genetic algorithms and bio-inspired optimization, support scheduling, resource leveling, and backlog prioritization in IT projects, enhancing operational efficiency and enabling dynamic adaptation to changing constraints (Taboada et al., 2023). Hybrid AI models, which integrate ML with fuzzy logic, optimization, or simulation techniques, improve predictive accuracy, supporting adaptive planning, risk management, and resource allocation across the IT project lifecycle (Adamantiadou et al., 2025).

In this context, explainable AI (XAI) frameworks and human-in-the-loop oversight are important for interpreting model outputs, ensuring transparency, and alignment with managerial judgment (Arora et al., 2025). They further enhance trust, allowing project managers to interpret AI generated insights and integrate them responsibly into decision-making processes (Okonkwo et al., 2025). Additionally, data privacy and ethical considerations are also important due to the sensitive nature of projects, clients, and data. AI methods use extensive datasets, which may include proprietary code, architectural designs, and performance metrics. Without appropriate safeguards, such methods may disclose confidential information or generate biased recommendations, potentially affecting resource allocation, task assignment, or risk management decisions (Aramali et al., 2025). Effective practices include data anonymization, secure AI pipelines, access control, and compliance with regulatory standards such as GDPR.

While the traditional AI techniques have significantly enhanced predictive and analytical capabilities in IT PM, they are in general data-driven and task-specific. The emergence of GenAI represents a deeper shift toward content generation, contextual reasoning, and interactive human-AI collaboration. GenAI provides additional capabilities by automating code generation, drafting project documentation, producing test scripts, and creating interactive reporting dashboards, allowing IT project teams to concentrate on cognitively demanding challenges and long-term strategic objectives. They can successfully address aspects related to team composition and task assignment, where AI evaluates skills, roles, and project requirements to propose optimized team structures and workflows (Chan, 2025). However, unlike traditional predictive AI systems, GenAI introduces distinct challenges, including hallucinated outputs, prompt sensitivity, contextual inconsistency, and increased risks of confidential data leakage through interactive use. These emerging risks, combined with the rapid integration of GenAI tools into IT PM, highlight the need for a systematic examination of their applications, limitations, and governance implications.

## Methodology

This systematic review provides a comprehensive examination of GenAI application for different IT PM process groups, emphasizing the identification of key challenges and future research directions. We adopted the Preferred Reporting Items for Systematic Reviews and Meta-Analyses (PRISMA) methodology (Page et al., 2021), which delivers explicit guidance for conducting systematic reviews and represents a widely used standard in review literature. PRISMA includes a structured set of reporting items for systematic reviews alongside a flow diagram depicting the review process. The PRISMA framework begins by requiring an unambiguous specification of the research goals that form the basis of the review:

- Identify research approaches that propose, design or develop GenAI models and techniques for IT PM process groups
- Analyze the integration of existing pre-trained GenAI models in PM tasks
- Assess the main activities of PM lifecycle tackled in the research literature through GenAI methods
- Identify challenges and highlight future research directions for GenAI potential in PM

Starting from these research questions, we have defined relevant keywords, split them into classes (5 keywords per class and combined them with the core keyword "project management" resulting in 25 query phrases:

*Master query: Keyword 1 [LLM, GPT, generative AI, chatbot, virtual assistant] + Keyword 2 [IT, Software, Agile, Engineering, Research and development] + [Project management]*

One important item in the PRISMA methodology is selecting the search databases. In this study, Web of Science (WoS) served as the primary scientific database (Li et al., 2018). WoS was selected for this PRISMA-based systematic review due to its comprehensive coverage of high-quality, peer-reviewed articles and its rigorous indexing standards. Moreover, it provides advanced and reproducible search functionalities, facilitating transparent search strategies in accordance with PRISMA guidelines. The "Topic" search criterion was used to identify publications in which the search terms appeared in the title, abstract, keywords or Keywords Plus, ensuring a relevant retrieval of studies aligned with the review objectives. The search and retrieving of research articles have been performed between in November 2025–January 2026. Figure 1 presents the PRISMA 2020 flow diagram for our study highlighting results processing through three main phases: identification, screening and inclusion. In the first phase 242 records were retrieved from the WoS database using 25 search phrases.

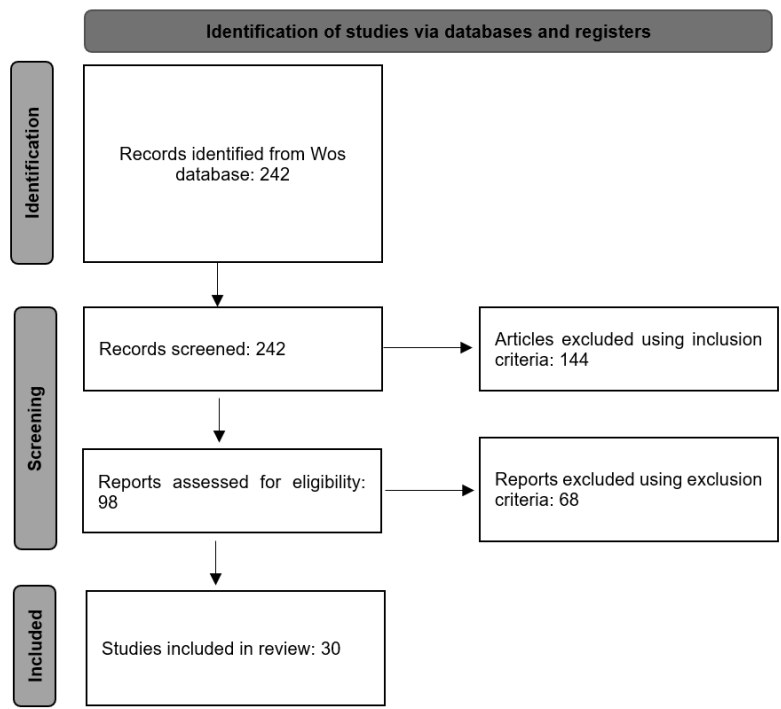

**Fig. 1. PRISMA 2000 flow diagram for selecting the review articles**

The subsequent Screening phase involved the application of predefined inclusion & exclusion criteria, which further filtered the records and reduced the number of eligible papers. The main *inclusion criteria* that narrowed the number of articles to 98 were: (i) publication year between 2021-2026, (ii) publication type (article or proceeding paper), (iii) English as publication language, (iv) restrict publishers (IEEE, Springer, Elsevier, MDPI, ACM, Taylor & Francis, Wiley, Nature, Sage and Oxford press), (v) research areas (computer science, engineering and Science Technology other topics) and (vi) WOS categories (Software engineering, Information systems, Computer science theory methods, AI, Computer science interdisciplinary and multidisciplinary applications). Next, we have performed a manual filtering of the retrieved papers using as *exclusion criteria*: (i) papers not related to GenAI, (ii) papers not related to IT software management and (iii) review articles marked as articles or proceedings papers. Additionally, we have used citation criteria for selecting most relevant approaches: minimum 10 citations for 2021-2023 papers and minimum 5 citations for 2024 papers. The final number of selected papers was 30.

The 30 reviewed papers were distributed across multiple publishers, with IEEE being the most represented (13 papers), followed by Elsevier and Springer (5 papers each), MDPI (3 papers), IEEE/ACM (2 papers), and Nature and ACM (1 paper each). The publications were evenly split between journal articles and conference proceedings (15 each). Among the journal publications indexed in Web of Science, 8 were classified as Q1 and 7 as Q2. The average number of citations per paper is 17, showing good quality for the selected articles. Figure 2 shows detailed statistics about the selected articles.

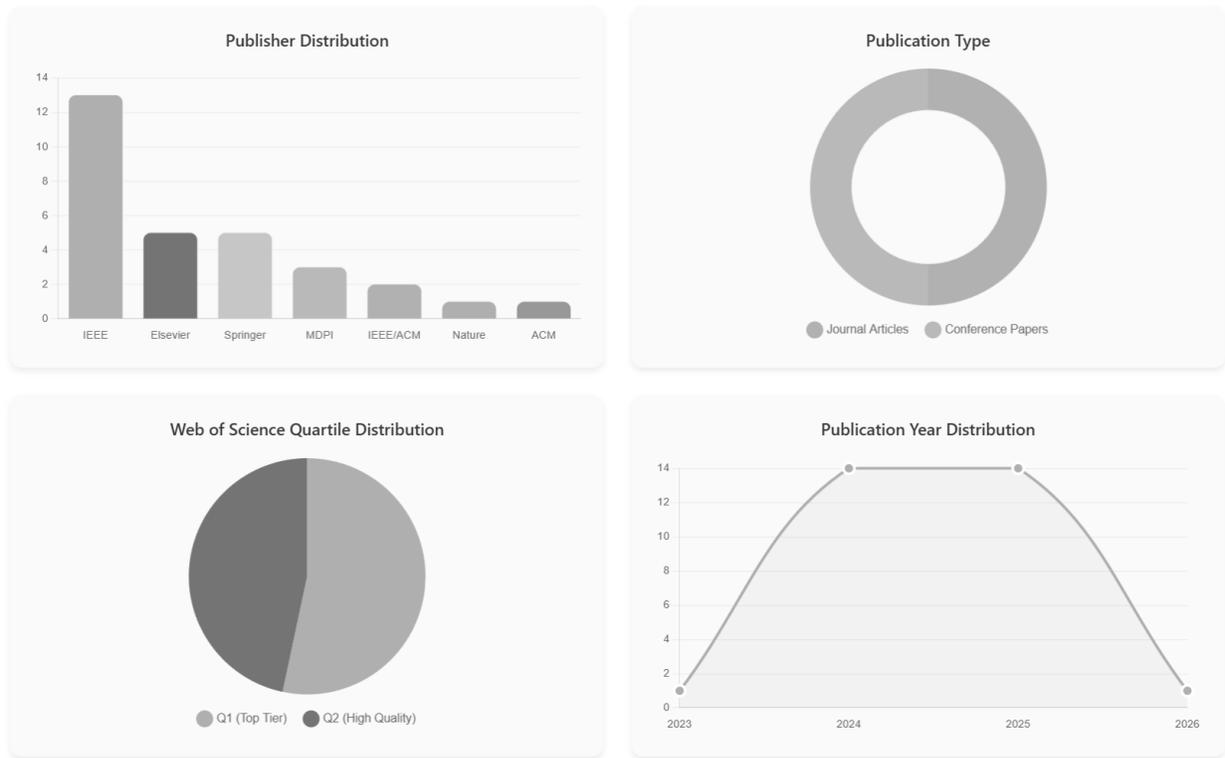

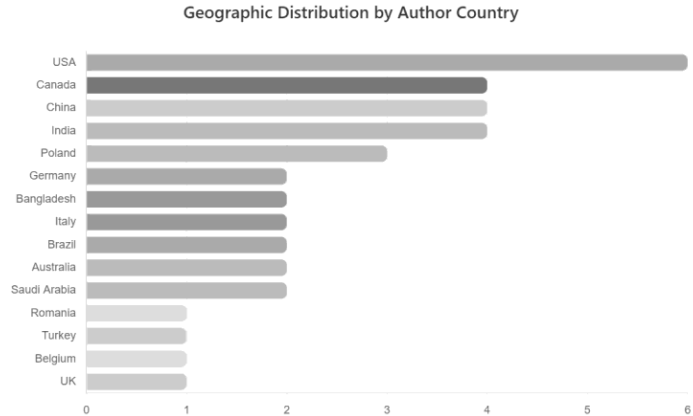

Fig. 2. Selected articles statistics

The analysis reveals a relatively balanced distribution between open access and subscription-based publications, with 14 papers (47%) available as open access and 16 papers (53%) published under subscription models. The geographic distribution of author affiliations demonstrates the global nature of research in this domain, with contributions from 20 different countries across multiple continents (see Figure 2). 7 of the 30 reviewed papers (23%) involved international collaborations with authors from multiple countries, reflecting the collaborative nature of the targeted domain.

## GenAI for IT PM process groups

There are few approaches in the selected studies that specifically address the **initiating** PM process group using GenAI techniques. (Mahbub et al., 2024) investigates how well GPT-4 with zero-shot strategy, can help in software requirements analysis for detecting defects such as ambiguities (where the wording can be interpreted in more than one way), inconsistencies (conflicting requirements) or incompleteness (missing elements or context). The model performed best at identifying incompleteness with a precision of 0.61 but it was also effective at spotting missing functional details. Similarly (Krishna et al., 2024) evaluates the efficacy of GPT-4 and CodeLlama in automating the creation and validation of software requirements specification documents. As input the authors use a university's student club management application specifications. ChatGPT outputs were significantly more precise and showed a higher resistance to hallucinations but resulted in information gaps and incomplete responses.

For the **planning** PM process group, most research approaches are focused on *using LLMs for document analysis*. (Rahman et al., 2024) investigate the use of LLMs for automating user story generation, a task closely related to requirements engineering and project backlog management. The paper employs gpt-4-turbo to generate user stories directly from natural language software descriptions. Rather than fine-tuning the models, the approach relies on a custom Refine and Thought prompt engineering technique and in-context learning, using structured prompts to guide the LLM toward producing user stories that follow standard agile templates (e.g., role–goal–benefit format). (Sarim et al., 2025) presents a framework for automating the decomposition of high-level software project descriptions into structured task flows. The researchers evaluated different state-of-the-art models, including Gemini 2.5 Pro, GPT-Omni, Grok 3, DeepSeek-R1, and LLaMA-3, to ensure the findings were not biased toward a specific architecture. To evaluate these models, five distinct prompting strategies were tested: Zero-Shot, Few-Shot, Chain-of-Thought (CoT), and a specialized ISO 21502-Guided approach that aligns task generation with international PM standards. (Lee et al., 2024) investigate the potential of fine-tuned LLMs to support knowledge management in engineering contexts, particularly for reviewing complex technical specification documents that are usually lengthy and challenging even for experienced engineers to assess comprehensively. The approach uses GPT3 - Davinci-002 and LLaMA2-13B LLMs together with a QLoRA fine-tunning technique to transform internal technical specifications into interactive question-and-answer interfaces that allow for faster, more intuitive information retrieval. The authors emphasize that while LLaMA2 shows stronger performance in most categories, the GPT-3 model performs better on more explicit requirements. (Dhruva et al., 2024) introduce a framework that integrates LLMs with Agile methodologies to enhance PM through data-driven automation. The system uses Google's PaLM LLM to process raw input (tickets or feature requests) and automatically generate task attributes (e.g. name, summary, priority, required team). It also uses a regression model trained on historical enterprise data to forecast the task's duration. Other researchers focus on integrating LLMs and GenAI technologies for *improving software architecture*. (Junior et al., 2025) examine the role of ChatGPT in supporting software architecture decision-making for microservice-based systems use-cases. The authors propose a prompt pattern sequence intended to guide architects through complex design choices. The analysis shows that different prompt patterns can produce good recommendations that consider project-specific technical and business constraints, but humans in the loop are still required before making any architectural choice. (Dhar et al., 2024) addresses the area of documenting Architectural Knowledge (AK) proposing an automated AKM system that leverages information extraction and GenAI (GPT-4 & Flan-T5) models to document AK and provide a query-based interface for architectural decision-making. The authors conducted an exploratory study focused specifically on generating architectural design decisions with promising results but highlighting the need for further research to achieve close to human performance. Finally, some techniques proposed in the literature for the planning process group deal with *generating tasks or specifications* using LLMs. (Barcaui et al., 2023) conduct a comparative analysis of applying GPT-4 mode in the context of developing a comprehensive project plan against a human PM. The research analyzes project plan components considering coherence, completeness, and structure of outputs produced by GPT-4 relative to those created by a human. It relies on prompt engineering to elicit structured project planning artifacts from the base LLM. The LLM demonstrates efficiency in generating broad coverage of planning items and consistency in format, while the human PM contributes with deeper contextual judgment and situation-specific nuances that AI alone currently lacks. The study emphasizes that generative LLMs can serve as assistive tools rather than function as standalone replacements. (Zhen et al., 2024) develop a task planning system based for creating a Work Breakdown Structure (WBS) with a three-layer architecture: agent layer performs task decomposition, the WBS layer aggregates sub-tasks by enforcing pre-encoded temporal, spatial, and resource constraints and the execution layer can be used to manage external components through specific communication protocols to perform actions in different scenarios. GPT 3.5/4 models are used for the agent layer to generate sub-tasks that are further processed in the WBS layer. Table 1 synthesizes the analyzed research for the planning process group.

**Table 1. Overview of GenAI research for the planning process group**

| Approach | GenAI Model | Main objective | Main GenAI usage | Techniques for GenAI |
|---|---|---|---|---|
| (Rahman et al., 2024) | gpt-4-turbo | Optimize user stories generation from textual descriptions | Document analysis | Fine tuning (e.g. QLoRA) |
| (Dhruva et al., 2024) | PaLM + ChatGPT | Optimize task scheduling | Document analysis (PaLM) + Synthetic data generation (ChatGPT) | Prompt engineering Retrieval Augmented Graphs |
| (Sarim et al., 2025) | LLaMA-3, Gemini 2.5 Pro, GPT-Omni, DeepSeek-R1 and Grok 3 | Translating raw software documentation into workflow representations | Document analysis | |
| (Junior et al., 2025) | ChatGPT | Support architects in micro-service architecture design | Generate technical and business architectural constraints | |
| (Barcaui et al., 2023) | GPT-4 | Generate detailed project plan | Document/specification generation | |
| (Zhen et al., 2024) | GPT3.5/4 | Generate detailed project plan | Tasks generation | |
| (Lee et al., 2024) | GPT3 and LLaMA2-13B | Technical documents and specification processing | Document analysis | |
| (Dhar et al., 2024) | GPT-4 & Flan-T5 | Extract architectural knowledge | Extract architectural knowledge from multiple sources | |

For the **executing process group**, LLM coding assistants are analyzed by (Slama and Lemire, 2025) to assess how they can affect developer productivity and what their performance and usability is in live coding environments. The authors analyze and classify a wide range of LLMs considering various criteria such as IDE integration, supported languages or cloud features. However, the article focusses on evaluating with more details only two widely used AI coding assistants, GitHub Copilot and TabNine through different metrics.

**Monitoring and controlling process group** runs usually concurrently with the executing process group to ensure the project stays on track and deals with performance, Key Performance Indicators (KPIs), and deviations management. A recent approach (Wysocki et al., 2026) targets issue tracking by analyzing data collected during monitoring and controlling process group. The proposed method is based on LLM for data augmentation with the general objective of improving an ML-based classification of recurring tasks (Transformer architecture with XGBoost classifiers). Different LLM models are employed through prompting strategies to generate synthetic data (ChatGPT-4-turbo), perform data reformation (Llama3.3) and project specific task generation (DeepSeek-R1-14B). Validation is carried out on data coming from six projects managed using Jira platform showing that LLMs can tackle the problem of imbalanced task categories and can bring improvements to the ML pipeline performance. (Colavito et al., 2025) tackles automated labeling of software issue reports (e.g. bug, feature, etc.), by analyzing how different LLMs perform in terms of response accuracy, quality, computational cost, and deployment feasibility. The authors benchmark 22 generative decoder-only LLMs and 2 encoder-only models (BERT based) that are evaluated on datasets with GitHub issues. The tested LLM models included in the benchmark are open-source LLMs (e.g. Mistral), code-specific (e.g. CodeLLaMA), and proprietary (e.g. using API calls with GPT-4). Datasets are used for training and testing the BERT-based models (RoBERTa and SETFIT) while for the LLMs prompt engineering is employed. The results show that generative decoder-only LLMs enable zero-shot issue classification without task-specific training, but their accuracy and consistency vary substantially across datasets while BERT-style encoder models show better performance and lower computational costs. (Gnieciak and Szandala, 2025) conducted a benchmark study comparing three major rule-based Static Analysis Security Testing (SAST) tools (SonarQube, CodeQL, and SnykCode) against three LLMs (GPT-4o/4.1, Mistral Large, and DeepSeek V3). Prompting is used for interacting with the LLMs through APIs by means of a custom developed tool, ProjectAnalyzer. Results show that LLMs outperform static analyzers in mean F-1 score due to superior recall and broader contextual reasoning but feature higher false-positive rates. However, running a full scan of a massive enterprise codebase via LLMs on premise is significantly more expensive and slower than SAST tools. (Akgül et al., 2025) argue that the complexity of AI integrated in software projects development often results in the accumulation of technical debt (TD), potentially affecting system reliability and long-term sustainability. They specifically address Data Debt (DD), a subset of TD that is unique to AI/ML systems where poor data quality creates hidden costs over time. 73 cases from 7 DD categories from real projects are analyzed by human experts and GPT-4o mini model with zero temperature by prompt engineering. Despite that GPT enhanced the overall process, some of the model's outputs were not, highlighting the importance of the human in the loop. Similarly, (Sheikhaei et al., 2024) research LLMs usage in supporting the identification and classification of self-admitted technical debt (SATD) (e.g. code, requirement, defect, etc.) within software repositories. The study compares a range of fine-tuned Flan-T5 family models with traditional convolutional neural network (CNN) classifiers to assess effectiveness in both detection and classification of SATD. Fine-tuned LLMs consistently outperform baselines in SATD identification while for classification, the study found that few-shot prompting can narrow the gap by incorporating task examples into prompts. (Badhon et al., 2025) propose a framework that combines a risk relationship insight component for building a knowledge graph to capture dependencies and cascading effects among risk factors and a risk factor influence analysis technique based on a Generative Adversarial Network (GAN). For the first technique, LLMs process written descriptions contained within the knowledge base to identify key concepts and map the semantic connections between them, enabling the capture of interdependence among risk factors using contextual embeddings and semantic associations. However, details about the LLM and associated tunings are not provided. (Pasam et al., 2025) present an approach to improving comment management in collaborative digital environments such as shared collaboration spaces in project development. The proposed system categorizes and prioritizes comments to help teams address the most critical feedback leveraging advanced NLP and ML methods, including transformer models (BERT and RoBERTa), which were fine-tuned for comment classification, as well as Hierarchical Capsule Networks (HcapsNet) and Hierarchical Attention Networks (HAN) to handle hierarchical comment structures. Additionally, the study explores the use of the GEMMA-2B LLM, demonstrating strong performance and precision through zero-shot and few-shot learning. (Chen et al., 2025) propose Meet2Mitigate, a framework that applies LLMs for project meeting analysis by combining multiple AI components into a hybrid system. It integrates automatic speech recognition to convert raw meeting audio into structured transcripts, which are then processed by an LLM for issue extraction and summarization. The authors employ a RAG strategy, allowing the LLM's outputs to take advantage of external best-practice knowledge bases when generating mitigation recommendations. The framework demonstrates how different LLMs (GPT, Llama, Mistral, Gemma and Qwen) can be embedded within a pipeline of supporting techniques to enable context-aware outputs. (Aggrawal & Magana, 2024) investigate the use of LLMs to support conflict management within team-based projects, implementing a three-phase procedure (learning, practice, and reflection) to enhance confidence and conflict resolution skills. The study uses ChatGPT 3.4 as interactive simulator during the practice phase to present and respond to conflict scenarios, enabling participants to engage in simulated conflict dialogues. The study does not integrate additional techniques such as RAG or fine-tuning. (Zhang et al., 2024) propose RustC4, a hybrid tool that combines LLMs with static program analysis to detect inconsistencies between code comments and implementation in Rust programs. The approach exploits LLMs' natural language understanding capabilities to interpret comment text and extract implicit design constraints, particularly those related to conditions and safety assumptions that are common in Rust's memory management model. The evaluation, conducted on a dataset derived from twelve large-scale real-world Rust projects, demonstrates the effectiveness of this LLM-assisted pipeline in identifying a substantial number of real comment–code inconsistencies. Table below summarizes the GenAI approaches for monitoring and controlling process groups.

**Table 2. Integrating GenAI techniques into the monitoring and controlling process group**

| Approach | GenAI Model | Main objective | Main GenAI usage | Techniques for GenAI |
|---|---|---|---|---|
| (Wysocki et al., 2026) | ChatGPT-4-turbo / Llama3.3 / DeepSeek-R1-14B | Issue tracking and classification | Synthetic data generation | Prompt engineering; integration with Transformer and XGBoost classifiers |
| (Chen et al., 2025) | GPT 3.5/4, Mistral, Llama 3.1/3.2, Gemma 2 and Qwen 2 | Project-related issues identification and solution suggestion out of meeting audio | Document (transcript) analysis | Prompt engineering + RAG; SBERT |
| (Sheikhaei et al., 2024) | Flan-T5 | Technical Debt analysis | Code analysis | Prompt engineering + context learning + fine tunning |
| (Colavito et al., 2025) | Mistral, CodeLLaMA, GPT-4, etc. | Automatic labeling of software issues | Zero-shot task classification | Prompt engineering |
| (Gnieciak and Szandala, 2025) | GPT-4o/4.1, Mistral Large, and DeepSeek V3 | Static Analysis Security Testing | code analysis | |
| (Akgül et al., 2025) | GPT-4o mini | Data Debt analysis | Document/specification analysis | |
| (Badhon et al., 2025) | N/A | Risk management | Extract concepts and semantic relations | |
| (Pasam et al., 2025) | GEMMA-2B | Optimize shared collaboration environments (intent classification) | Document analysis | |
| (Aggrawal, & Magana, 2024) | ChatGPT 3.5 | Teams conflict management | Scenario analysis | |
| (Zhang et al., 2024) | ChatGPT | detect inconsistencies between code comments and implementation | Code analysis | |

For the **closing process group** most of the selected approaches target maintenance aspects using LLMs mostly for code analysis. (Yang et al., 2025) propose ChatDL, a LLM based solution to address the challenge of software maintenance specifically focusing on how to efficiently locate software defects for Industrial Internet of Things (IIoT) use-cases. It combines ChatGPT, a multisource gate network, an enhancement of ColBERT model, and a reinforcement learning algorithm with mean reciprocal rank (MRR) as reward function. Evaluation metrics assessment show that ChatDL outperformed existing models (e.g. FBL-BERT) on publicly available Java datasets. (Gandhi et al., 2025) propose a framework that combines algorithmic analysis (to detect code divergence) with LLMs based Multi-Agent Systems (MAS). The system automatically identifies which commits need to be ported, analyzes the contextual gap between branches, and generates adaptive code modifications. It uses OpenAI o1-preview LLM for semantic code clones' detection and code generation. For the development of the MAS system, Autogen platform is used together with different LLMs such as GPT4, Gemini or Llama. However, the approach does not deeply explore the security implications of using LLMs with external APIs on proprietary, closed-source codebases where data privacy is a priority. (Zhao et al., 2024) addresses the efficiency of LLMs in fixing software bugs, specifically by introducing the concept of Design Rationales. The authors propose DRCodePilot, a framework that functions through a structured pipeline. First, it performs rationale extraction, identifying the design intent behind successful patches by leveraging data from platforms like GitHub and Jira. To address GPT-4's lack of project-wide context, the framework incorporates a self-reflective mechanism, in which the model reviews and refines its output using a suggested rationale as guidance.

However, there are approaches that can be classified as **multiple** PM process groups (see Table 3). (Alliata et al., 2025) explore how LLMs can support routine Agile PM tasks, particularly those typically related to a Scrum Master. Claude 3 Sonnet, Microsoft Copilot and ChatGPT 4.5 models are evaluated in several scenarios using Rouge, Meteor and BERTScore metrics to assess the accuracy and quality of resulted data. More specifically the approach focusses on two use-cases using prompt engineering: generating specific synthetic data during planning PM process group (e.g. creating user stories from epics) and status reports in the monitoring and controlling PM process group (e.g. interpreting burndown chart). The authors argue that LLMs can interpret charts and produce narrative descriptions, but results vary with context that is provided in prompts. LLMs also did well to create sound user stories from epics this allowing to optimize backlog refinement and requirements documentation processes.

Table 3. Approaches targeting multiple PM process groups

| Approach | GenAI Model | Main objective | Main GenAI usage | Techniques for GenAI |
|---|---|---|---|---|
| (Alliata et al., 2025) | Claude 3 Sonnet, ChatGPT 4.5, Microsoft Copilot | User stories and status reports creation | Produce narrative descriptions | Prompt engineering |
| (Bala et al., 2025) | ChatGPT 4.0 | improve different PM process groups documentation | Document generation | |
| (Sainio et al., 2023) | ChatGPT | Automatizing and managing project tasks | Document analysis for management tasks | |
| (Alam et al., 2025) | oasst-sft-1-pythia-12b | Generate insights into software specifications | AI chatbot for document | |

|  |  |  | processing and analysis |  |
| --- | --- | --- | --- | --- |
| (Faruqui et al., 2024) | GPT-J | SDLC Analysis and management | Document analysis | Context learning; Fine tunning |
| (Cinkusz et al., 2025) | GPT-4/3.5 | Optimize user tasks | Take decisions for different PM roles | Prompt engineering and parameters configuration (LangChain); Agentic AI (CogniSim) |

In (Cinkusz et al., 2025) the authors study how LLMs can be integrated in the Scaled Agile Framework (SAFe) for optimizing PM tasks. The approach focuses on creating cognitive agents for emulating human roles in Agile methodologies (e.g. Product Owner, System Architect, etc.) and simulate evaluation scenarios using CogniSim framework. The integrated LLMs are OpenAI's GPT-4 and GPT-3.5 models by taking advantage of the LangChain framework. Each agent is represented by an LLM instance that specializes in its role using prompt engineering and parameters configuration (e.g. temperature). The results show that the proposed cognitive agents can improve PM metrics such as task completion time and communication coherence. However, how the LLM agents perform reasoning isn't deeply explored. (Alam et al., 2025) propose an AI Chatbot for project managers that can analyze PDF documents data streams and generate insights with the goal of improving decision making and even replacing projects managers in certain situations. The chatbot is built around OpenAssistant LLM (oasst-sft-1-pythia-12b) which is enhanced with techniques such as tokenization, character text splitter, or lemmatization for data processing steps. The solution combines as input PDF files with specific prompts. Compared with other PDF chatbots such as ChatPDF it has good values for metrics such as cosine similarity of or semantic similarity score (>80%) having as main advantage the specialization for the PM domain. ChatGPT 4.0 is used by (Bala et al., 2025) to process and improve documentation used in the PM chain. The authors tackle two aspects, text quality and its adjustment for the right person in the PM process groups. Prompt engineering is used to generate different document versions and evaluation is done through metrics such as text characteristics (e.g. number of paragraphs) and readability (e.g. Coleman-Liau Index). Results have shown that the generated versions featured improved readability and understandability, however the method was evaluated only on three documentation samples, and no supplementary tailoring was done to the LLM model. Prompt Engineering using ChatGPT as a strategic tool to mitigate PM challenges and automate routine managerial tasks is seen as a fundamental technique by (Sainio et al., 2023) to address issues in Agile such as role ambiguity, communication gaps, and stakeholder management. The researchers evaluated ChatGPT's performance across various PM tasks by designing prompt patterns for different critical areas of PM. The findings emphasize that while ChatGPT's knowledge base is generally reliable for PM for automating specific sub-tasks, LLMs could not replace human project managers, especially in phases requiring complex emotional or ethical decision-making. (Faruqui et al., 2024) present AI-Analyst, a framework that leverages LLMs to assist in Software Development Life Cycle (SDLC) analysis with the goal of optimizing business costs. It interprets and synthesizes textual artefacts across the SDLC, including requirements, design descriptions, test reports, and maintenance logs. The framework adapts GPT-J to analyze different phases of SDLC based on a transfer learning-based approach and using a dataset from comprehensive SDLC documentation.

The analysis of the included studies reveals a clear dominance of OpenAI's GPT/ChatGPT family, which reflects its widespread accessibility and strong general-purpose capabilities that make it the default choice for researchers exploring GenAI in PM. Alternatives such as LLaMA, Gemini and Mistral suggest a growing but still secondary interest in LLM models. Regarding techniques, most studies rely solely on prompt engineering, indicating that most research remains at an exploratory or proof-of-concept stage rather than pursuing deeper model adaptation. More advanced approaches such as RAG, fine-tuning, or agentic AI frameworks appear in few studies, highlighting a significant gap in the literature and an opportunity for future research to move beyond surface-level prompting toward more domain-specific integration of GenAI in PM contexts.

## GenAI Integration into PM Platforms

As GenAI technologies mature, they can be further embedded directly into PM platforms, transforming how work is planned, executed, and reviewed in IT and software contexts. These integrations mark a shift from standalone AI tools toward AI that is tightly coupled with everyday work environments used by organizations, such as Jira, Confluence, or Microsoft 365 Copilot. Research into how GenAI is used within project settings highlighted that many view AI features in these tools as assistants, co-pilots, or supportive agents rather than replacements for human project managers. For example, (Assalaarachchi et al., 2025) reports that GenAI features are commonly used to automate routine tasks—such as generating project plans and reports, summarizing meetings, and composing follow-up emails—and to support agile practices without disrupting established workflows. However, in the last years significant efforts have been put into integrating new AI features in the major PM available commercial or open-source platforms.

**Commercial platforms** such as *Atlassian* (Jira, Bitbucket or Confluence) use GenAI functionalities to assist users in drafting and summarizing content, generating test plans and project documentation, or managing communications (Atlassian, 2025). These capabilities are designed to speed up content creation, reduce cognitive load, and improve team collaboration by providing context-aware suggestions directly within the tools that teams already use daily. *Microsoft 365 Copilot* integrates LLMs into Microsoft 365 applications to assist users in drafting documents, generating reports, answering questions, and retrieving information across multiple tools (Microsoft Copilot, 2026). These integrations bring GenAI support into project status updates, stakeholder communications, and collaborative planning, while also creating predictive insights and contextual summaries that would traditionally require manual effort. *monday.com* has embedded AI throughout its Work OS platform through its AI assistant and automation engine (monday.com, 2025) to generate task descriptions, suggest workflow automations, and produce summaries of board activity and project progress. It also offers predictive timeline adjustments when task dependencies shift. *AsanaAI* adds AI-driven features aimed at improving prioritization and execution to Asana's work management platform (AsanaAI, 2026). It offers smart summaries of project status, AI-generated goals and milestones, and automatic identification of at-risk tasks based on workload and deadline patterns. Its Smart Workflow feature allows users to build automation rules using natural language.

In addition to commercial platforms, **open-source and GitHub-based projects** are increasingly approaching GenAI integration into PM workflows. For instance, the ProjectManagementLLM offers pretrained LLMs tailored to PM tasks, enabling automated task suggestions and report generation (ProjectManagementLLM, 2025). AutoScrum automates the Scrum project planning process using LLMs by generating a comprehensive set of agile artefacts including user stories, features, sprint goals, acceptance criteria, tasks and requirements thus automating much of the work involved in sprint preparation (AutoScrum, 2026). It is designed as a planning assistant rather than a replacement for humans. Similarly, Scrum Sage Zen Edition is a GPT-powered assistant built on the ChatGPT platform, specifically designed to support Scrum practitioners in applying agile principles (Scrum Sage, 2026). It functions as an interactive coaching and facilitation tool, helping teams and Scrum Masters work through sprint ceremonies, refining their backlog practices, and navigating common agile challenges through conversational guidance. Kanban-MCP is a middleware layer that connects LLMs to Planka, an open-source Kanban board application, via the Model Context Protocol (MCP) (Kanban-MCP, 2026). It enables AI assistants such as Claude to directly interact with Kanban board data managing tasks and labels, adding comments, and tracking time spent on individual work items. Agentic Project Management (APM) platform tackles integrating agents in the PM processes (Agentic Project Management, 2026) through an AI workflow framework that coordinates a team of specialized AI agents such as project manager agent, developer agent, etc. to manage complex projects through structured multi-agent workflows. One of its core design goals is addressing the context window limitations in LLMs, using memory and context retention techniques to ensure that productive AI-assisted work can be sustained across longer sessions without loss of important project state.

In summary, GenAI's integration into both commercial and open-source PM platforms signals a significant evolution in how IT project work is supported. These deployments bring AI assistance into core project processes, making GenAI a practical component of everyday project lifecycle activities even though most of the integrations are based on API-based third-party cloud models. Indeed, the computing power required to specialize and deploy personalized and privacy-aware LLMs into the premises of an organization remains a challenge. Thus, continued research is required to define how such tools will be adopted, how they can influence managerial roles and responsibilities, and how issues like privacy and ethics can be handled in enterprise settings.

## Perspective of AI agents in PM

Our study shows that GenAI is increasingly integrated into both commercial and open-source IT project management platforms, automating routine tasks, generating project documentation, assisting in coding and testing, and supporting decision-making processes. Additionally, there is a growing trend of community-driven solutions that leverage GenAI to optimize IT PM processes while remaining accessible for experimentation and adaptation in diverse project contexts. Also, across the reviewed studies, GenAI has demonstrated potential to enhance productivity, enable faster and more informed decisions, and support human–AI collaboration throughout various project process groups.

However, as AI becomes increasingly integrated into project management, we identify and discuss three promising directions that are emerging:

1. *Process group specialized AI agents*, providing targeted support for planning, executing, and reporting.
2. *AI agents that replicate and simulate project management roles*, acting as virtual team members such as project managers, risk managers, and scrum masters.
3. *Hybrid of process group-specific agents informed by role-based simulations*, creating a network of AI collaborators that can operate in parallel while human managers oversee the final decisions.

They all aim to enhance decision-making while keeping humans actively in the loop having the potential of transforming PM from reactive oversight to proactive orchestration.

**In the first direction**, project process group specialized AI agents assist in planning by analyzing historical data and resource needs, streamline execution through real-time progress tracking, and enhance reporting by generating insights automatically. They will enable more efficient and informed decision-making while reducing the workload on human managers. *Project initiating AI agents* could autonomously gather and analyze stakeholder inputs, organizational documents, and market context to generate preliminary project charters, define scope boundaries, and identify key risks and success criteria. Specific artifacts that AI agents could generate or manage at this stage include project charters, business case documents, stakeholder registers, feasibility assessment reports, initial risk logs, and scope definition statements. Dedicated *planning AI agents* can design project roadmaps, allocate resources, and generate sprint backlogs based on project objectives and team capacities. They will integrate historical project data, budget constraints, and dependency mappings. These agents could simultaneously optimize scheduling, risk mitigation strategies, and communication plans, while offering the outputs for human review, refinement and approval. Artifacts include draft WBS, Gantt charts, resource allocation matrices, risk registers, sprint backlog documents, cost estimation reports, etc. The *AI agents specialized for executing* can act as mediators across teams, synthesizing information, coordinating dependencies, and facilitating stakeholder communication. They could autonomously assign work based on real-time availability and skill profiles, generate code, documentation, and test scripts, and resolve minor blockers without human intervention, escalating only those issues that require managerial judgment. Key artifacts managed or generated during this process group include code repositories, technical documentation, user stories, issue tracking records, test plans and automated test scripts. *Project monitoring and controlling AI agents* could continuously ingest performance data, detect deviations from baselines, diagnose root causes, and suggest mitigation strategies in real time shifting the PM from reactive problem-solving toward anticipatory, intelligence-driven control. Artifacts include performance dashboards, earned value analysis reports, risk reassessment logs, change request documents, or sprint retrospective summaries. The *closing AI agents* could autonomously compile project performance reports, draft compliance documentation for audits or regulatory needs, and extract and categorize lessons learned from communication logs and issue trackers. This would transform closing from an often-neglected administrative task into a structured, AI-driven knowledge capture process that continuously enriches the organization's institutional knowledge base and improves future project planning. Specific artifacts include final project completion reports, lessons learned repositories, post-implementation review documents, and updated organizational knowledge bases.

The **second direction** focuses on training AI agents that replicate and simulate specific project management roles, acting as virtual team members with specialized responsibilities such as project sponsors, business analysts, scrum masters, or technical leads. Each role-based agent monitors relevant aspects of the project, analyzes data from a shared knowledge layer, and makes recommendations or takes automated actions within its domain. A *Project Sponsor AI agent* (i.e. Product Owner in Agile methodologies such as Scrum) oversees strategic alignment and executive decision-making, ensuring the project is organizationally supported. Can be trained on business cases, investment portfolios, governance frameworks, and historical executive decision logs to evaluate project viability and highlight critical issues requiring top-level intervention. *A Business Analyst AI agent* will assure the connection between stakeholders and technical teams by identifying business needs, defining and monitoring requirements. Past project charters, requirement documents, user stories, and change requests can be used for specializing in and training this agent. The *Technical Lead AI agent* will define the architectural view of the system and will coordinate technical decisions to ensure that deliverables are feasible and aligned with design standards. Specific context can be given through technical documentation and architecture records, code review logs, and past implementation reports. The *Quality Assurance Lead AI agent* ensures standards are met by analyzing performance metrics and past quality reports by using data from past quality reports, testing logs or performance benchmarks. The *Risk Manager AI agent* proactively identifies threats and opportunities, proposing mitigation and contingency plans with the help of historical risk registers and scenario simulations using historical risk registers, incident reports, and scenario simulations. Finally, *Stakeholder Management AI agent* monitors engagement, sentiment, and stakeholder needs, leveraging feedback data, communication logs, and sentiment analysis to foster collaboration and support. Collectively, these AI agents should learn from historical and real-time project data, shared knowledge bases, continuously adapting to organizational practices and industry standards.

**The third direction** combines process group-specific AI agents with role-based AI agents, creating a hybrid network of AI collaborators that operate in parallel while human project managers maintain oversight (see Fig. 3). In this model, agents retain their specialized responsibilities but also may simulate project management roles like project manager, risk manager, or scrum master, allowing them to provide context-aware recommendations across the project lifecycle. The hybrid approach enables AI agents to anticipate potential conflicts, optimize resource allocation, and proactively suggest adjustments before issues escalate.

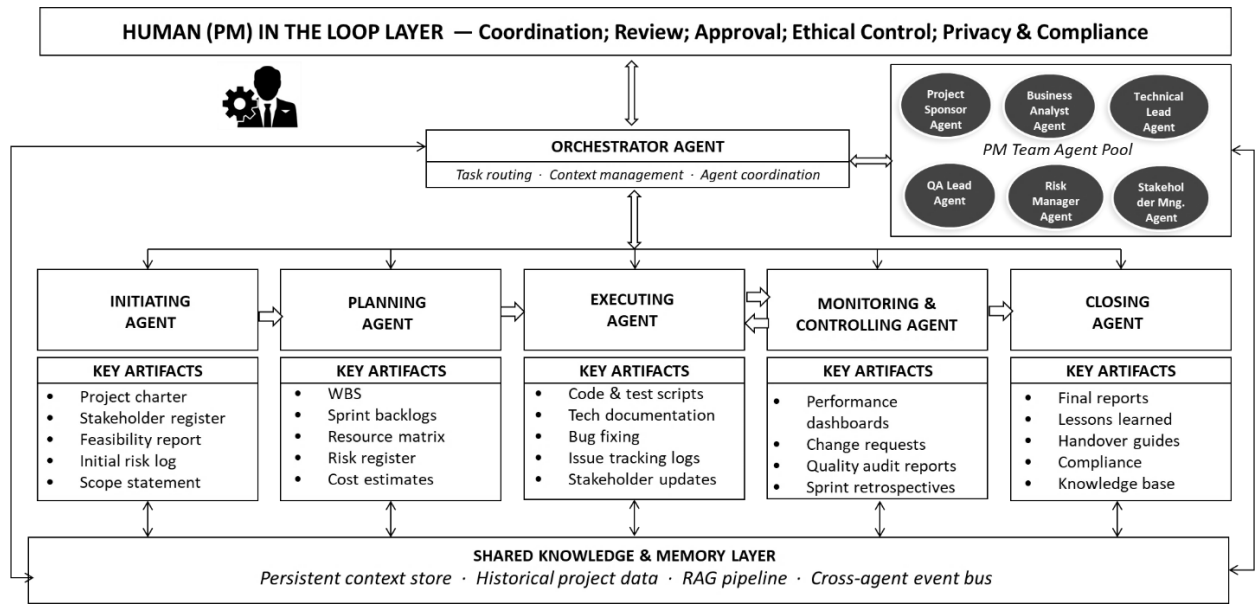

Fig. 3. Agentic AI Collaborative Architecture for PM

A **central component** of the three envisioned agentic directions (see Fig. 3) is the **Orchestrator Agent**, which functions as the coordinator of the multi-agent system. This agent will be the support tool for the Project Manager managing the overall project plan, tracking progress across all domains, and ensuring coherence between all other role-based agents. It will be responsible for decomposing high-level project objectives into subtasks, dynamically routing these tasks to the appropriate specialized agents based on context, priority, and project process group, and managing the flow of information across the system. It maintains a persistent representation of the overall project state, ensuring that each agent operates with consistent and up-to-date context, and resolves conflicts or dependencies that arise when multiple agents interact with shared resources or overlapping concerns. The Orchestrator Agent also will be the primary interface between the human project manager and the PM AI agent's ecosystem, receiving human directives, translating them into machine-executable instructions, and offering support for decisions that need human oversight. In this context, a shared knowledge layer needs to be defined as the backbone for AI-driven project management, providing a central repository of project data, insights, and organizational rules that all AI agents, either process group-specific or role-based can access and update. This layer needs to store structured data such as project plans, timelines, resource allocations, and risk registers, as well as unstructured information like reports, meeting notes, and communications. Built on a semantic or knowledge-graph framework, it will map relationships between tasks, resources, risks, and stakeholders, enabling AI agents to infer dependencies, anticipate conflicts, and propose solutions across project process groups. Agents could communicate with each other through this layer, flagging issues, sharing updates, and coordinating actions, while all interactions are logged to ensure traceability and accountability.

To realize this vision for the future of agentic project management, **significant research, technical, and organizational challenges must be addressed**. First challenge is data quality and integration for AI models training. Project data are often fragmented across tools (e.g., scheduling software, ERP systems, communication platforms) and stored in heterogeneous formats. Inconsistent, incomplete, or biased historical data can significantly affect model performance (Li et al., 2025). Building a unified, high-quality shared knowledge layer requires standardization protocols and secure integration pipelines. Additionally, to make things even more challenging, effective AI models training requires aggregated data from multiple projects across different contexts, teams, and levels of complexity to better capture variability and reduce biases (Abdelzaher et al., 2025). Moreover, projects evolve and industries change over time. AI agents trained in outdated data may produce suboptimal or misleading recommendations. Continuous monitoring and data ingestion, retraining pipelines are necessary to maintain agents' relevance and accuracy. Second challenge lies in interoperability and coordination among AI agents. In a multi-agent system, process group-specific and role-based agents must exchange information without creating conflicts, redundancies, or feedback loops (Sapkota et al., 2026). Designing clear communication protocols, conflict-resolution mechanisms, and hierarchical or consensus-based decision models is important to ensure coherence. Third, challenge lies in model transparency and explainability (Hughes, Mavi et al., 2025). Project decisions often have financial, contractual, and strategic consequences. If AI agents provide recommendations without interpretable reasoning, human managers may resist adoption. Organizations will need to establish rules for agent autonomy, decision traceability, and human control. Ethical and regulatory considerations such as bias mitigation, transparency, and secure handling of sensitive data will be essential to ensure trust and adoption of multi-agent PM ecosystems. Explainable AI mechanisms are necessary to maintain accountability. In this context security, privacy, and compliance are major concerns (Papagiannidis et al., 2025). Project data frequently contains sensitive financial, contractual, and personal information. A shared knowledge layer accessible by multiple agents requires strong cybersecurity controls, role-based access management, encryption, and compliance with regulatory standards. Finally, human-AI collaboration and role clarity must be carefully managed to make clear the responsibilities between humans and machines (Göndöcs et al., 2025). Clear governance structures are required to define decision authority, escalation procedures, and approval thresholds, ensuring that humans remain accountable for strategic and ethical judgments. Teams may perceive AI agents as surveillance tools or job replacements rather than collaborative assistants. Successful adoption will require transparent communication, training programs, and gradual integration strategies that position AI agents as collaborators rather than automation.

## Conclusion

This review examined the evolving role of GenAI, in the context of IT project management. By synthesizing recent academic literature, and analyzing commercial and open-source platforms, the paper highlights how GenAI technologies are reshaping core project management processes across the project lifecycle. The analysis shows that GenAI is increasingly embedded into widely used PM platforms, as well as into experimental and open-source solutions developed within the software engineering community. These integrations enable automation of documentation, backlog refinement, reporting, and communication tasks, thereby reducing administrative overhead and supporting agile and adaptive project practices. At the same time, the findings suggest that GenAI is predominantly positioned as a supportive assistant or co-pilot, augmenting rather than replacing the role of the human project manager. Additionally, challenges related to data privacy, governance, transparency, and human–AI collaboration remain insufficiently explored in the current literature. Overall, the findings show the need for further empirical research and integrative reviews that examine GenAI's impact on project performance, managerial decision-making, and professional roles in IT project management. As GenAI continues to evolve and become more deeply embedded in PM platforms, new research initiatives will focus on building and integrating project management process group specialized AI agents, agents that replicate and simulate project management roles, and finally a hybrid of process group-specific agents informed by role-based simulations, creating a network of AI collaborators that can operate in parallel while human managers oversee the final decisions.

## References

Abdelzaher, T., Hu, Y., Kara, D., et al. (2025). The bottlenecks of AI: Challenges for embedded and real-time research in a data-centric age. *Real-Time Systems, 61*, 185–236. https://doi.org/10.1007/s11241-025-09452-w

Abdulghafour, M., & Chirchir, B. (2024). Challenges of integrating artificial intelligence in software project planning: A systematic literature review. *Digital, 3*, 555–571.

Adamantiadou, D. S., & Tsironis, L. (2025). Leveraging artificial intelligence in project management: A systematic review of applications, challenges, and future directions. *Computers, 14*(66).

Agentic Project Management: Autonomous AI for agile workflows. (2026). GitHub. https://github.com/sdi2200262/agentic-project-management

Aggrawal, S., & Magana, A. J. (2024). Teamwork conflict management training and conflict resolution practice via large language models. *Future Internet, 16*(5), 177. https://doi.org/10.3390/fi16050177

Akgül, N. Y., Temizel, T. T., Top, Ö. Ö., & Akman, P. D. (2025). Aligning data debt with AI-integrated software project lifecycle processes: A standard-based mapping approach. In *2025 IEEE/ACM International Conference on Technical Debt (TechDebt)* (pp. 1–11). IEEE. https://doi.org/10.1109/TechDebt66644.2025.00008

Alam, K., Bhuiyan, M. H., Islam, M. S., Chowdhury, A. H., Bhuiyan, Z. A., & Ahmmed, S. (2025). Co-pilot for project managers: Developing a PDF-driven AI chatbot for facilitating project management. *IEEE Access, 13*, 43079–43096. https://doi.org/10.1109/ACCESS.2025.3548519

Alharthi, M., Abid, M. A., & Ahmed, S. H. (2023). Artificial intelligence in project management: A systematic review. *Systems, 13*(10), 913.

Alliata, Z., Singhal, T., & Bozagiu, A. M. (2024). The AI Scrum Master: Using large language models (LLMs) to automate agile project management tasks. In *Agile Processes in Software Engineering and Extreme Programming – Workshops (XP 2024)* (pp. 1–15). Springer. https://doi.org/10.1007/978-3-031-72781-8_12

Almeida, P. M., Fernandes, G., & Santos, J. M. R. C. A. (2025). Artificial intelligence tools for project management: A knowledge-based perspective. *Project Leadership and Society, 6*, 100196. https://doi.org/10.1016/j.plas.2025.100196

Aramali, V., Cho, N., Pande, F., Al-Mhdawi, M. K. S., Ojiako, U., & Qazi, A. (2025). Generative AI in project management: Impacts on corporate values, employee perceptions, and organizational practices. *Project Leadership and Society, 6*, 100191.

Arora, L., et al. (2025). Explainable artificial intelligence techniques for software development lifecycle: A phase-specific survey. In *IEEE COMPSAC 2025*.

AsanaAI. (2026). Get started with Asana AI. https://help.asana.com/s/article/get-started-with-asana-ai

Assalaarachchi, L. I., Masood, Z., Hoda, R., & Grundy, J. (2025). Generative AI for software project management: Insights from a review of software practitioner literature. *IEEE Software*. https://doi.org/10.1109/MS.2025.3619936

Atlassian. (2025, January 14). Harness the power of AI editing across Atlassian products. https://www.atlassian.com/blog/announcements/ai-editing-atlassian-products

AutoScrum: AI-assisted agile project management. (2026). GitHub. https://github.com/autoscrum/autoscrum

Badhon, B., Chakrabortty, R. K., Anavatti, S. G., & Vanhoucke, M. (2025). A multi-module explainable artificial intelligence framework for project risk management. *Engineering Applications of Artificial Intelligence, 148*, 110427. https://doi.org/10.1016/j.engappai.2025.110427

Bala, S., Sahling, K., Haase, J., & Mendling, J. (2024). ChatGPT for tailoring software documentation for managers and developers. In *XP 2024 Workshops* (pp. 1–15). Springer. https://doi.org/10.1007/978-3-031-72781-8_11

Barcaui, A., & Monat, A. (2023). Who is better in project planning? Generative artificial intelligence or project managers? *Project Leadership and Society, 4*, 100101. https://doi.org/10.1016/j.plas.2023.100101

Chan, J., & Li, Y. (2025). Enhancing team diversity with generative AI: A novel project management framework. *arXiv*. https://arxiv.org/abs/2502.05181

Chen, G., et al. (2025). Meet2Mitigate: An LLM-powered framework for real-time issue identification and mitigation. *Advanced Engineering Informatics, 64*, 103068. https://doi.org/10.1016/j.aei.2024.103068

Cinkusz, K., Chudziak, J. A., & Niewiadomska-Szynkiewicz, E. (2025). Cognitive agents powered by large language models for agile software project management. *Electronics, 14*(1), 87. https://doi.org/10.3390/electronics14010087

Colavito, G., Lanubile, F., & Novielli, N. (2025). Benchmarking large language models for automated labeling. *Information and Software Technology, 184*, 107758. https://doi.org/10.1016/j.infsof.2025.107758

Dhar, R., Vaidhyanathan, K., & Varma, V. (2024). Leveraging generative AI for architecture knowledge management. In *ICSA-C 2024* (pp. 163–166). IEEE. https://doi.org/10.1109/ICSA-C63560.2024.00034

Dhruva, G., Shettigar, I., Parthasarthy, S., & Sapna, V. M. (2024). Agile project management using large language models. In *ICITIIT 2024* (pp. 1–6). IEEE. https://doi.org/10.1109/ICITIIT61487.2024.10580873

Dong, H., Dacre, N., Baxter, D., & Ceylan, S. (2024). What is agile project management? *Project Management Journal, 55*(6), 668–688.

Faruqui, N., et al. (2024). AI-Analyst: An AI-assisted SDLC analysis framework for business cost optimization. *IEEE Access, 12*, 195188–195203. https://doi.org/10.1109/ACCESS.2024.3519423

Gandhi, A., et al. (2025). Automated codebase reconciliation using large language models. In *IEEE Forge 2025* (pp. 1–11). IEEE. https://doi.org/10.1109/Forge66646.2025.00011

Gnieciak, D., & Szandala, T. (2025). Large language models versus static code analysis tools. *IEEE Access, 13*, 198410–198422. https://doi.org/10.1109/ACCESS.2025.3635168

Göndöcs, D., Horváth, S., & Dörfler, V. (2025). Human-AI hybrid performance dynamics. *International Journal of Human-Computer Studies, 205*, 103622. https://doi.org/10.1016/j.ijhcs.2025.103622

Hughes, L., Dwivedi, Y. K., Li, K., Appanderanda, M., Al-Bashrawi, M. A., & Chae, I. (2025). AI agents and agentic systems. *Journal of Global Information Technology Management, 28*(3), 175–185.

Hughes, L., Mavi R.K., et al. (2025). Impact of artificial intelligence on project management. *Journal of Innovation & Knowledge, 10*(5), 100772. https://doi.org/10.1016/j.jik.2025.100772

Junior, J. J. M., Melegati, J., & Guerra, E. (2025). Applying a prompt pattern sequence. In *ESOCC 2025* (pp. 1–10). Springer. https://doi.org/10.1007/978-3-031-84617-5_2

Kanban-MCP: AI-supported Kanban project management. (2026). GitHub. https://github.com/bradrisse/kanban-mcp

Karnouskos, S. (2024). The relevance of large language models for project management. *IEEE Open Journal of the Industrial Electronics Society, 5*, 758–768. https://doi.org/10.1109/OJIES.2024.3412222

Khalil, M., Bravo, A., Vieira, D., & Carvalho, M. M. (2025). Mapping the AI landscape in project management context. *Systems, 13*(10). https://doi.org/10.3390/systems13100913

Krishna, M., Gaur, B., Verma, A., & Jalote, P. (2024). Using LLMs in software requirements specifications. In *RE 2024* (pp. 475–483). IEEE. https://doi.org/10.1109/RE59067.2024.00056

Lee, J., Kim, H., & Park, Y. (2025). Large language models for agile project management. *arXiv*. https://arxiv.org/abs/2509.26014

Lee, J., Jung, W., & Baek, S. (2024). In-house knowledge management using a large language model. *Applied Sciences, 14*(5), 2096. https://doi.org/10.3390/app14052096

Li, K., Rollins, J., & Yan, E. (2018). Web of Science use in published research. *Scientometrics, 115*, 1–20.

Li, X., Cheng, Y., Møller, C., & Lee, J. (2025). Data issues in industrial AI systems. *Computers in Industry, 173*, 104361. https://doi.org/10.1016/j.compind.2025.104361

Mahbub, T., et al. (2024). Can GPT-4 aid in detecting ambiguities? *IEEE Access, 12*, 171972–171992. https://doi.org/10.1109/ACCESS.2024.3464242

Microsoft. (2026). Microsoft 365 Copilot overview. https://learn.microsoft.com/en-us/copilot/microsoft-365/microsoft-365-copilot-overview

monday.com. (2025). monday.com AI features. https://monday.com/p/press-release/monday-com-unveils-platform-wide-ai-shift-the-work-execution-era-arrives/

Müller, R., Locatelli, G., Holzmann, V., Nilsson, M., & Sagay, T. (2024). Artificial intelligence and project management. *Project Management Journal, 55*(1), 9–15.

Nenni, M. E., et al. (2025). How artificial intelligence will transform project management. *Management Review Quarterly, 75*, 1669–1716. https://doi.org/10.1007/s11301-024-00418-z

Okonkwo, R., Folorunso, A., Ogundipe, F., & Tettey, C. Y. (2025). Explainable artificial intelligence through human-AI collaboration. *Computer Science & IT Research Journal, 6*(5), 333–354.

Page, M. J., et al. (2021). The PRISMA 2020 statement. *Systematic Reviews, 10*, 89.

Papagiannidis, E., Mikalef, P., & Conboy, K. (2025). Responsible artificial intelligence governance. *Journal of Strategic Information Systems, 34*(2), 101885. https://doi.org/10.1016/j.jsis.2024.101885

Pasam, V. K., Pati, S., & Hernandez, C. T. (2025). AI-powered comment triage. In *SAC 2025* (pp. 971–979). ACM.

PMBOK Guide. (2026). *A guide to the project management body of knowledge (PMBOK® Guide)* (8th ed.). https://www.pmi.org/standards/pmbok

ProjectManagementLLM: Generative AI for project management tasks. (2025) https://huggingface.co/ai-in-projectmanagement/ProjectManagementLLM

Rahman, T., et al. (2024). Take loads off your developers. In *ICSME 2024* (pp. 791–801). IEEE. https://doi.org/10.1109/ICSME58944.2024.00082

Rose, K. H. (2013). PMBOK Guide Fifth Edition. *Project Management Journal, 44*(3), e1.

Sainio, K., Abrahamsson, P., & Ahtee, T. (2023). Prompt patterns for agile software project managers. In *ICSOB 2023* (pp. 1–10). Springer.

Salimimoghadam, S., et al. (2025). The rise of artificial intelligence in project management. *Buildings, 15*(7), 1130. https://doi.org/10.3390/buildings15071130

Sapkota, R., Roumeliotis, K. I., & Karkee, M. (2026). AI agents vs. agentic AI. *Information Fusion, 126*, 103599. https://doi.org/10.1016/j.inffus.2025.103599

Sarim, M., et al. (2025). Generating reliable software project task flows. *Scientific Reports, 15*, 35194. https://doi.org/10.1038/s41598-025-19170-9

Scrum Sage Zen Edition Version 2. (2026). ChatGPT Labs. https://chatgpt.com/g/g-PSusbDrAK-scrum-sage-zen-edition-version-2

Sheikhaei, M. S., et al. (2024). An empirical study on large language models for SATD identification. *Empirical Software Engineering, 29*, 159.

Singh, P., Choudhary, R., & Zhou, L. (2025). AI-based project planning using generative models. *arXiv*. https://arxiv.org/abs/2510.10887

Slama, F., & Lemire, D. (2025). Enhancing developer productivity. In *IDS 2025* (pp. 39–45). IEEE. https://doi.org/10.1109/IDS66066.2025.00011

Su, X., & Ayob, A. H. (2025). Artificial intelligence in project success. *Information, 16*, 682. https://doi.org/10.3390/info16080682

Taboada, I., Daneshpajouh, A., Toledo, N., & de Vass, T. (2023). Artificial intelligence enabled project management. *Applied Sciences, 13*, 5014.

Wang, H., Zhao, L., & Chen, Y. (2025). Artificial intelligence in project management. *Journal of Project Analytics, 5*, 15–28.

Wysocki, W., & Ochodek, M. (2026). Leveraging LLM-based data augmentation. *Journal of Systems and Software, 231*, 112641. https://doi.org/10.1016/j.jss.2025.112641

Yang, H., Zhou, Y., Liang, T., & Kuang, L. (2025). ChatDL: An LLM-based defect localization approach. *IEEE Internet of Things Journal, 12*(16), 32333–32343. https://doi.org/10.1109/JIOT.2025.3531512

Zhang, Y., Liu, Z., Feng, Y., & Xu, B. (2024). Leveraging LLM to detect Rust code comment inconsistency. In *ASE 2024* (pp. 356–366). IEEE/ACM.

Zhao, J., Yang, D., Zhang, L., Lian, X., Yang, Z., & Liu, F. (2024). Enhancing automated program repair. In *ASE 2024* (pp. 1706–1718). IEEE/ACM.

Zhen, Y., et al. (2024). LLM-Project: Automated engineering task planning. In *CYBER 2024* (pp. 605–610). IEEE. https://doi.org/10.1109/CYBER63482.2024.10749328